\documentclass[12pt,a4paper]{article}
\pdfoutput=1

\usepackage[utf8]{inputenc}
\usepackage[T1]{fontenc}

\usepackage{multirow}
\usepackage{amsmath}
\usepackage{color}
\usepackage{amsfonts}
\usepackage{amssymb}
\usepackage{graphicx}
\usepackage{geometry}
\usepackage{amssymb,epsfig,subfigure}
\usepackage{hyperref}
\usepackage{url}
\usepackage{comment}
\usepackage[font=footnotesize]{caption}
\usepackage{slashed}

\makeatletter
\renewcommand\section{\@startsection {section}{1}{\z@}%
                                 {-3.5ex \@plus -1ex \@minus -.2ex}%nn
                                   {2.3ex \@plus.2ex}%
                                   {\normalfont\large\bfseries}}
\renewcommand\subsection{\@startsection{subsection}{2}{\z@}%
                                   {-3.25ex\@plus -1ex \@minus -.2ex}%
                                     {1.5ex \@plus .2ex}%
                                     {\normalfont\bfseries}}
\renewcommand\subsubsection{\@startsection{subsubsection}{3}{\z@}%
                                   {-3.25ex\@plus -1ex \@minus -.2ex}%
                                     {1.5ex \@plus .2ex}%
                                     {\normalfont\itshape}}
\makeatother

\def\pplogo{\vbox{\kern-\headheight\kern -29pt
\halign{##&##\hfil\cr&{\ppnumber}\cr\rule{0pt}{2.5ex}&\ppdate\cr}}}
\makeatletter
\def\ps@firstpage{\ps@empty \def\@oddhead{\hss\pplogo}%
  \let\@evenhead\@oddhead % in case an article starts on a left-hand page
}%      The only change in \maketitle is \thispagestyle{firstpage} instead of 
\def\maketitle{\par
 \begingroup
 \def\thefootnote{\fnsymbol{footnote}}
 \def\@makefnmark{\hbox{$^{\@thefnmark}$\hss}}
 \if@twocolumn
 \twocolumn[\@maketitle]
 \else \newpage
 \global\@topnum\z@ \@maketitle \fi\thispagestyle{firstpage}\@thanks
 \endgroup
 \setcounter{footnote}{0}
 \let\maketitle\relax
 \let\@maketitle\relax
 \gdef\@thanks{}\gdef\@author{}\gdef\@title{}\let\thanks\relax}
\makeatother

\numberwithin{equation}{section}

\newcommand\eea{\end{eqnarray}}
\newcommand\bea{\begin{eqnarray}}

\def\beq{\begin{equation}}
\def\eeq{\end{equation}}

\newcommand{\be}{\begin{equation}}
\newcommand{\ee}{\end{equation}}
\newcommand{\ba}{\begin{align}}
\newcommand{\ea}{\end{align}}
\newcommand{\bg}{\begin{gather}}
\newcommand{\eg}{\end{gather}}
\newcommand{\bseq}{\begin{subequations}}
\newcommand{\eseq}{\end{subequations}}

\renewcommand{\ln}{\mathop{\rm ln}\nolimits}

\renewcommand{\t}{\tilde}

\newcommand{\coment}[1]{}

\begin{document}
\setcounter{page}0
\def\ppnumber{\vbox{\baselineskip14pt
%\hbox{hep-th/0000000}
}}
\def\ppdate{
%\footnotesize{SU/ITP-14/XX}
} \date{}

\author{Jeremías Aguilera-Damia$^1$, Mario Sol\'is$^{2, 3}$, Gonzalo Torroba$^{2, 3}$\\
[7mm] \\
{\normalsize \it $^1$  Physique Th\'eorique et Math\'ematique and International Solvay Institutes}\\
{\normalsize \it Universit\'e Libre de Bruxelles; C.P. 231, 1050 Brussels, Belgium}\\
{\normalsize \it $^2$ Instituto Balseiro, UNCuyo and CNEA}\\
{\normalsize \it $^3$Centro At\'omico Bariloche and CONICET}\\
{\normalsize \it S.C. de Bariloche, R\'io Negro, R8402AGP, Argentina}\\
}

\bigskip
\title{\bf  Nonrelativistic Dirac fermions on the torus
\vskip 0.5cm}
\maketitle

\begin{abstract}
Two dimensional conformal field theories have been extensively studied in the past. When considered on the torus, they are strongly constrained by modular invariance. However, introducing relevant deformations or chemical potentials pushes these theories away from criticality, where many of their aspects are still poorly understood. In this note we make a step towards filling this gap, by analyzing the theory of a Dirac fermion on the torus, deformed by a mass term and a chemical potential for the particle number symmetry. The theory breaks conformal and Lorentz invariance, and we study its spectrum and partition function. We also focus on two limits that are interesting on their own right: a massless relativistic fermion with nonzero chemical potential (a simple model for CFTs at finite density), and nonrelativistic Schrodinger fermions (of relevance in condensed matter systems). Taking inspiration from recent developments in massive modular forms, we obtain a representation of the torus free energy based on Fourier-transforming over a twisted boundary condition. This dual representation fullfills many properties analogous to modular invariance in CFTs. In particular, we use this result to derive Cardy-like formulas for the high energy density of states of these theories. 
\end{abstract}
\bigskip

\newpage

\tableofcontents

\vskip 1cm

%%%%%%%%%%%%%%%%%%%%%%%%%%%%%%%%%%%%%%%%%
%%%%%%%%%%%%%%%%%%%%%%%%%%%%%%%%%%%%%%%%%
%%%%%%%%%%%%%%%%%%%%%%%%%%%%%%%%%%%%%%%%%
%%%%%%%%%%%%%%%%%%%%%%%%%%%%%%%%%%%%%%%%%
\section{Introduction and summary}\label{sec:intro}

Two-dimensional conformal field theories (CFTs) on the torus have been extensively studied during the last decades. Modular invariance, arising from discrete reparametrizations of the torus, is a powerful tool that has led to deep insights on CFTs. Among many important results based on modular transformations, we can highlight the Cardy formula for the density of states \cite{Cardy:1986ie}, the modular bootstrap initiated by \cite{Hellerman:2009bu}, and relations to 3d gravity \cite{Strominger:1997eq, Maloney:2007ud}. On the other hand, massive theories have been comparatively less studied on the torus; for instance, modular properties of massive fermions have only been fully clarified very recently \cite{Berg:2019jhh, Downing:2023uuc, Berg:2023pca}. Moreover, very little is known about nonrelativistic quantum field theories (QFTs) on the torus, see e.g. \cite{Gonzalez:2011nz, Castro:2015uaa, Melnikov:2018fhb, Chen:2019hbj}. The relevance of nonrelativistic QFTs to condensed matter physics and more specifically to quantum criticality \cite{Fradkin:2013sab} provides an important motivation for analyzing this type of models. These theories break modular invariance explicitly, so traditional methods from CFT do not work, and need to be modified.
In this work we seek to improve this situation.

Nonrelativistic QFT appears to be too general to allow for a fruitful analysis. But it is useful to recall that many general lessons have been discovered by studying free models \cite{DiFrancesco:1997nk}. Following this lead, here we will focus our attention on a free Dirac fermion in two space-time dimensions, with nonzero mass $m$ and chemical potential $\mu_F$,
\be\label{eq:action0}
S= \int d^2x\, \bar \psi(\gamma^\mu \partial_\mu- \gamma^2 \mu_F+m) \psi\,,
\ee
where $\mu=2$ is the euclidean time direction.
For $m=\mu_F=0$, this is a simple example of a conformal field theory, with a rich interplay between modular transformations, spin structures, and duality \cite{DiFrancesco:1997nk}. Turning on a nonzero mass, but still with $\mu_F=0$, gives a Lorentz invariant field theory with a nontrivial but calculable renormalization group flow. When placed on the torus, the mass parameter explicitly breaks modular invariance, but this can be restored if one allows for a rescaling of the mass. This is a very useful perspective, which leads to interesting mathematical generalizations of elliptic functions and modular forms \cite{Berg:2019jhh}. On the other hand, once the chemical potential is nonzero, Lorentz invariance is explicitly broken. This gives rise to a nonzero charge density and, by Pauli exclusion principle, the vacuum state describes a free Fermi liquid \cite{Shankar:1993pf}. Our goal will be to analyze the theory with nonzero $m$ and $\mu_F$. Besides its clear physical interest, we will find that this also leads to mathematical generalizations of different modular structures.

We will organize our analysis in three different regimes. First, in Sec. \ref{sec:massless} we will focus on $m \to 0$ with finite $\mu_F$. This is the simplest situation for our purpose, and we fill find that the partition function admits an expression in terms of certain complexifications of elliptic functions that describe the conformal field theory. This is a simple model for exploring CFTs with nonzero chemical potential, and we will also make contact with previous works on this subject.
Next, in Sec. \ref{sec:massive} we will analyze the case of generic and nonzero $(m, \mu_F)$; even though the chemical potential breaks Lorentz invariance, we will find that its effect can be understood using generalizations of the massive modular forms recently introduced in \cite{Berg:2019jhh}. Finally, in Sec. \ref{sec:nonrel} we will consider a nonrelativistic Schrodinger fermion, which arises from the large mass limit of a Dirac fermion with $m-\mu_F$ fixed.

One of our main results will be expressions for partition functions as double sums, that arise from a particular Fourier representation of the free energy. This will generalize the recent results of \cite{Berg:2019jhh, Downing:2023uuc, Berg:2023pca} to the nonrelativistic realm. Even though these theories break modular invariance explicitly, in a sense this double-sum representation appears as a natural generalization of modular covariance. A key application of our methods will be to determine the high energy density of states. This will provide an extension of the Cardy formula to nonrelativistic QFT.

%%%%%%%%%%%%%%%%%%%%%%%%%%%%%%%%%%%%%%%%%%%%%%%
%%%%%%%%%%%%%%%%%%%%%%%%%%%%%%%%%%%%%%%%%%%%%%%
%%%%%%%%%%%%%%%%%%%%%%%%%%%%%%%%%%%%%%%%%%%%%%%
%%%%%%%%%%%%%%%%%%%%%%%%%%%%%%%%%%%%%%%%%%%%%%%
\section{Massless Dirac fermion}\label{sec:massless}

In this section we consider the massless case, $m=0$, with nonzero chemical potential. After reviewing the classical action, we compute the spectrum and partition function. Although Lorentz invariance is explicitly broken by the chemical potential, we find that the results can be written in terms of complexifications of elliptic functions that describe the theory with $\mu_F=0$. The last part of the section is devoted to presenting an explicitly modular covariant expression for the free energy. Using this we will obtain a Cardy formula valid at finite density.

\subsection{Action}

We will consider the euclidean field theory defined on a torus, specified by the identifications
\bea
(x^1, x^2) &\simeq & (x^1+L, x^2) \nonumber\\
(x^1,\,x^2)& \simeq &(x^1+L \tau_1, \, x^2+L \tau_2)\,,
\eea
where $x^2$ is the Euclidean time direction, $L$ fixes the size, and $\tau= \tau_1+i \tau_2$ is the complex structure for the torus. We choose the Dirac representation
\be\label{eq:Dirac}
\gamma^1 = i\left(\begin{matrix}
0 & -1 \\ 1 & 0
\end{matrix} \right)\;,\;\gamma^2 = \left(\begin{matrix}
0 & 1 \\ 1 & 0
\end{matrix} \right)\,.
\ee
 It will also be useful to work with coordinates $(\sigma^1, \sigma^2)$ in a standard square  $0 \le \sigma^i \le 1$, related to the previous ones by
 \be
 \sigma^1 + \tau \sigma^2 = \frac{1}{L}(x^1+ i x^2)\,.
 \ee

In components,
\be
\psi=\left(\begin{matrix} \psi_1 \\ \psi_2 \end{matrix} \right)\,,
\ee
the action (\ref{eq:action0}) for $m=0$ becomes
\be
S= \int d^2x\,\left[ \psi_1^\dag (\partial_2+i \partial_1-\mu_F) \psi_1+\psi_2^\dag (\partial_2-i \partial_1-\mu_F) \psi_2 \right]\,.
\ee
In order to exhibit the explicit dependence on the torus parameters $L$ and $\tau$, we change to the $\sigma$ coordinates, finding
\be
S= L\,\int d^2 \sigma\, \left[\psi_1^\dag(- \bar \tau \partial_{\sigma^1}+ \partial_{\sigma^2}- L \tau_2 \mu_F) \psi_1 +\psi_2^\dag(-  \tau \partial_{\sigma^1}+ \partial_{\sigma^2}- L \tau_2 \mu_F) \psi_2\right]\,.
\ee

\subsection{Spectrum}

Let us introduce the chiral Dirac operators
\be\label{eq:defD}
D=-  \tau \partial_{\sigma^1}+ \partial_{\sigma^2}- L \tau_2 \mu_F\;,\;\bar D=- \bar \tau \partial_{\sigma^1}+ \partial_{\sigma^2}- L \tau_2 \mu_F
\ee
and compute the eigenvalues and eigenvectors,
\be
D \psi_n = \lambda_n \psi_n\,.
\ee

We are interested in both periodic and antiperiodic boundary conditions in the spatial and thermal circles. They can be written as
\bea\label{eq:bc}
\psi^{(ab)}(\sigma^1+1,\sigma^2)&=&e^{2\pi i a}\psi^{(ab)}(\sigma^1,\sigma^2)\nonumber\\
\psi^{(ab)}(\sigma^1,\sigma^2+1)&=&e^{2\pi i b}\psi^{(ab)}(\sigma^1,\sigma^2)\,,
\eea
where $a, b=0,\,1/ 2$.

The eigenfunctions and eigenvalues of $D$ are
\bea\label{eq:eigenf}
\psi_n^{(ab)}(\sigma)& =& e^{2\pi i\left[(n_1+a) \sigma^1+(n_2+b) \sigma^2 \right]}\;,\;n_i \in \mathbb Z\nonumber\\
\lambda_n^{(ab)}&=&-2\pi i \tau (n_1+a)+2\pi i (n_2+b)- L \tau_2 \mu\,.
\eea
Note that due to the chemical potential, in general there is no zero eigenvalue. The eigenvalues $\bar \lambda$ of $\bar D$ are obtained by replacing $\tau \to \bar \tau$.

\subsection{Torus partition function}

The fermion path integral gives the formal expression for the torus partition function,
\be\label{eq:Zstart}
Z_{ab}= \det(D_{ab}) \det(\bar D_{ab}) = \prod_n\,\lambda_n^{(ab)}\,\bar \lambda_n^{(ab)}\,.
\ee
This is formal since it has to be regularized and we employ the usual zeta function regularization. The zero point energy is also ambiguous, and we will discuss it shortly.

Let us focus on the holomorphic part of the partition function,\footnote{For the other chirality, we have $\det(\bar D)= \det(D)^*$, so that the partition function is real, as it should be. This can be shown by redefining  $n_1+a \to -(n_1+a)$ and $n_2+b \to - (n_2+b)$ in the expression for $\det(\bar D)$. Equivalently, one can rename $\psi_2 \to \psi_2^\dag$ in the original action.}
\be
\det(D_{ab}) = \prod_{n_1,n_2=-\infty}^\infty \left[n_2+b-\tau(n_1+a)+i \frac{L \tau_2}{2\pi} \mu_F \right]\,.
 \ee
To make contact with the standard representation in terms of elliptic functions, it is useful to define
\be
q=e^{2\pi i \tau}
\ee
and
\be
y= e^{2\pi i z}\;,\;z=b+i \frac{L \tau_2}{2\pi} \mu_F\,.
\ee
As we already see here, the chemical potential complexifies the thermal boundary condition, $b \to z$.

Using the product formula
\be\label{eq:prod}
\prod_{n=-\infty}^\infty (n+\alpha)= e^{i \pi \alpha}- e^{-i\pi \alpha}\,,
\ee
the product over $n_2$ gives
\bea
\det(D_{ab}) &=&\prod_{n_1=-\infty}^\infty (e^{i \pi z}q^{-\frac{n_1+a}{2}}-e^{-i \pi z}q^{\frac{n_1+a}{2}}) \nonumber\\
&=& 2i \sin(\pi z-\pi \tau a)\, \prod_{n_1=1}^\infty\,(-1) q^{-n_1} (1-y q^{n_1-a})(1-y^{-1} q^{n_1+a})\,.
\eea
Using zeta function regularization,\footnote{Recall that
\be
\zeta(s) = \sum_{n=1}^\infty n^{-s}\;,\; \zeta(-1) =- \frac{1}{12}\;,\;\zeta(0)=- \frac{1}{2}\,.
\ee
In particular,
\be
\prod_{n=1}^\infty q^{-n} = e^{-\log q \,\sum_{n=1}^\infty n} = e^{-\log q\,\zeta(-1)}=q^{1/12}\,.
\ee
}
we arrive at
\be
\det(D_{ab})= -2\,\sin(\pi z-\pi \tau a)\, q^{1/12}\prod_{n=1}^\infty\,(1-y q^{n-a})(1-y^{-1} q^{n+a})\,.
\ee

Let us consider first the case of periodic boundary conditions on the spatial circle, $a=0$. Recalling the definitions (we follow the conventions in Ch. 10 of \cite{DiFrancesco:1997nk})
\be
\theta_1(z|\tau)= -i y^{1/2} q^{1/8}\,\prod_{n=1}^\infty(1-q^n)\, \prod_{n=0}^\infty (1-y q^{n+1}) (1-y^{-1} q^n)\,,
\ee
and
\be
\eta(\tau) = q^{1/24}\prod_{n=1}^\infty(1-q^n)\,,
\ee  
after some algebraic manipulations we obtain
\be
\det(D_{a=0,b})= -\frac{\theta_1(z|\tau)}{\eta(\tau)} \,.
\ee
The partition function then reads
\be
Z_{a=0,b}= \frac{|\theta_1(z|\tau)|^2}{|\eta(\tau)|^2}\,.
\ee
This agrees with the CFT result, up to the complexification $b \to z$ due to the chemical potential. In particular, the partition function $Z_{a=0, b=0}$ no longer vanishes at finite density.
 
For antiperiodic boundary conditions, $a=1/2$, the triple product expression reads
\bea
\theta_4(z|\tau)&=& \prod_{n=1}^\infty (1-q^n)\,\prod_{n=0}^\infty(1-y q^{n+1/2}) (1-y^{-1} q^{n+1/2})\nonumber\\
&=&q^{-1/24}\eta(\tau) \,\prod_{n=0}^\infty(1-y q^{n+1/2}) (1-y^{-1} q^{n+1/2})\,,
\eea
hence leading to
\be
\det(D_{a=\frac{1}{2},b})=i y^{1/2} q^{-1/8}\,\frac{\theta_4(z|\tau)}{\eta(\tau)}\,.
\ee
When $\mu_F=0$, this contains an extra factor of $q^{-1/8}$ compared to the operatorial calculation \cite{DiFrancesco:1997nk}. The difference is due to the ambiguity in the zero point energy. Taking the convention that the zero point energy vanishes for a CFT in flat space amounts to subtracting this extra factor, and hence
\be
Z_{a=\frac{1}{2},b}=\frac{|\theta_4(z|  \tau)|^2}{|\eta(\tau)|^2} \,.
\ee

In particular, for the spin structure $a=b=1/2$ (AA) we can use the shift property that relates $\theta_4$ to $\theta_3$ to find
\be\label{eq:Zaa}
Z_{a=\frac{1}{2},b=\frac{1}{2}}=\left |\frac{\theta_3(i \frac{L \tau_2}{2\pi} \mu_F|\tau)}{\eta(\tau)}\right|^2 \,.
\ee
Similar effects were found before in the context of entanglement entropy calculations at finite density \cite{Kim:2017xoh}.

%%%%%%%%%%%%%%%%%%%%%%%%%%%%%%%%%%%%%%%%%%%%%%%
%%%%%%%%%%%%%%%%%%%%%%%%%%%%%%%%%%%%%%%%%%%%%%%
\subsection{Modular transformations}\label{subsec:modtransf}

We will now analyze the effects of modular transformations with nonvanishing chemical potential. Recall that a modular transformation is a discrete diffeomorphism,
\be\label{eq:diff}
\left( \begin{matrix} \sigma^1 \\ \sigma^2 \end{matrix} \right)=\left( \begin{matrix} d && b \\ c && a \end{matrix} \right)\left( \begin{matrix} \sigma'^1 \\ \sigma'^2 \end{matrix} \right)\;,\;a,b,c,d\,\in \mathbb Z\;,\;ad-bc=1\,.
\ee
This corresponds to an equivalent choice of torus lattice. Furthermore, the metric changes as
\bea\label{eq:metric-transf}
ds^2 &=& |d\sigma^1+\tau d\sigma^2|^2 \nonumber\\
&=& |c\tau +d|^2 \left|d\sigma'^1+ \frac{a\tau+b}{c \tau+d}d\sigma'^2 \right|^2\,,
\eea
so the modular transformed complex parameter is
\be
\tau'=\frac{a\tau+b}{c \tau+d}\,.
\ee

A QFT placed on a torus need not be modular invariant. First, non-marginal interactions will not be invariant under the overall Weyl rescaling in (\ref{eq:metric-transf}). Also, if the theory is not Lorentz invariant, in general it will not be invariant under the coordinate transformation (\ref{eq:diff}), which mixes the space and time directions. The chemical potential combines these two effects. It is useful to understand the breaking of modular invariance by performing a spurion analysis, where we view $\mu_F$ as an expectation value of a field that is now allowed to transform. 

The chemical potential can be obtained by turning on an external gauge field $A_i$ that couples to the $U(1)$ electric current. More specifically, we need a nonzero time component, $\mu_F = i A_2$, but let us first explore the modular transformation of a general $A_i$. Since this is a vector of mass dimension 1, we can read off its transformation rule $A_i \to A_i'$ by changing the space-time coordinates $x \to x'$ according to (\ref{eq:diff}), and then equating $A_i dx^i = A_i' dx'^i$. A short calculation gives
\bea
A_1' &=& (c \tau_1 +d) A_1+ c \tau_2 A_2 \nonumber\\
A_2'&=&(c \tau_1 +d) A_2- c \tau_2 A_1\,.
\eea
For what follows, it is also convenient to write this transformation rule in terms of $A \equiv A_1 + i A_2$ and $\bar A= A_1-i A_2$:
\be\label{eq:Atransf}
A'=(c \bar \tau+d) A\;,\;\bar A'=(c \tau +d) \bar A\,.
\ee
In particular, a chemical potential $A_1$ in one frame gives rise to both a chemical potential $A_2'$ and a current source $A_1'$ in a different frame. 

Now, coupling the theory to the above gauge field and computing the eigenvalues \eqref{eq:eigenf} one verifies that $2\pi z= iL\tau_2 A$, hence leading to the following transformation under \eqref{eq:diff}
\be\label{eq:ztransf}
z\to \frac{z}{c\tau +d}\,.
\ee  

Let us apply this general discussion to the free fermion partition function (\ref{eq:Zaa}), focusing on the case of a modular S-transformation
\be
\tau' = - \frac{1}{\tau}\,.
\ee
This will also play a role in the analysis below of the density of states at high energies.
We can use the Jacobi identity that gives the S-transformation law
\be
\theta_3\left(\frac{z}{\tau}\Big |- \frac{1}{\tau}\right)= (-i \tau)^{1/2} e^{i \pi z^2/\tau}\,\theta_3(z|\tau)\,,
\ee
together with
\be
\eta(-1/\tau) = (-i \tau)^{1/2} \eta(\tau)\,.
\ee
Hence
\bea\label{eq:dualZ}
Z_{a=\frac{1}{2},b=\frac{1}{2}}(z, \tau) &=&\left |\frac{\theta_3(z|\tau)}{\eta(\tau)}\right|^2=|e^{-i \pi z^2/\tau}|^2 \left|\frac{\theta_3\left(\frac{z}{\tau}\Big |- \frac{1}{\tau}\right)}{\eta(-1/\tau)} \right|^2 \nonumber\\
&=&|e^{-i \pi z^2/\tau}|^2\,Z_{a=\frac{1}{2},b=\frac{1}{2}}(z/\tau, -1/\tau)\,.
\eea

We notice that the transformation in the argument of the elliptic function reproduces (\ref{eq:ztransf}). However,
we find that the partition function is not exactly invariant, but rather changes by the overall factor $|e^{-i \pi z^2/\tau}|^2$. Specializing to $\tau= i \tau_2$, we recognize this as a ground-state energy contribution due to the chemical potential, since $|e^{-i \pi z^2/\tau}|^2=e^{\frac{1}{2\pi}\tau_2 (L \mu_F)^2}$. This result is consistent with previous analyses of CFTs in finite-charge sectors \cite{Benjamin:2016fhe, Dyer:2017rul}, where this term was seen to arise from the path integral measure.

In passing, let us briefly comment on the relation of the above partition functions to the bosonic theory attained by the usual bosonization map. On general grounds, a fermionic theory is characterized by its coupling to the spin structure through the non-anomalous $(-1)^F$ ${\mathbb Z}_2$ symmetry. From this perspective, a choice of spin structure is equivalent to a particular configuration of the background field for $(-1)^F$.  Gauging this symmetry naturally leads to a theory which is independent of the spin structure, hence bosonic.\footnote{One might also perform the gauging with discrete torsion, namely by stacking the fermionic theory with an invertible fermionic topological order $(-1)^{{\rm Arf}(\rho)}$, generically leading to a different bosonic theory (see for instance \cite{Ji:2019ugf}).} More precisely  
\be\label{eq:bosonic}
Z^{bos} = \frac12 \sum_{\rho}Z^{ferm}[\rho]=\frac12\sum_{a,b=0,1/2} Z^{ferm}_{ab}
\ee
For the case of a single Dirac fermion CFT, the corresponding bosonic theory is the $c=1$ compact scalar $\phi\sim\phi+2\pi$ at radius $R=\sqrt{2}$ (in conventions for which the self-dual radius is $R=1$). Furthermore, mass and chemical potential deformations are mapped respectively to a potential $\cos(\phi)$ and a chemical potential for the winding (topological) symmetry on the bosonic side. It would be interesting to further exploit the techniques presented in this paper to study this class of bosonic theories.

%%%%%%%%%%%%%%%%%%%%%%%%%%%%%%%%%%%%%%%%%%%%%%%
%%%%%%%%%%%%%%%%%%%%%%%%%%%%%%%%%%%%%%%%%%%%%%%
\subsection{A dual representation}\label{subsec:dual}

Recently, \cite{Berg:2019jhh} proposed a massive generalization of modular forms, and this was then used by \cite{Downing:2023uuc, Berg:2023pca} to give an explicitly modular covariant representation of the partition function for a massive Dirac fermion. The approach is based on the idea of Fourier-transforming the free energy with respect to the fermion boundary conditions $(a, b)$. We will find that this is a powerful method for understanding the high temperature limit, and that its utility extends to cases where modular invariance does not play a useful role, as we will see in Sec. \ref{sec:nonrel}.

The first step is to observe that the partition function is invariant under $(a,b) \to (a+1, b+1)$, as can be seen from (\ref{eq:bc}). Therefore, we can Fourier-transform the free energy,
\be\label{eq:dualZ1}
\log Z_{ab}= \sum_{l, r \in \mathbb Z}\,f_{l,r}\, e^{-2\pi i (l a + r b)}
\ee
with coefficients
\be
f_{l,r}= \int_0^1 da \int_0^1 db\,\log Z_{ab}\, e^{2\pi i (l a + r b)}\,.
\ee
We will call this the `dual' representation.
Mathematically, this method appears for instance in the Eisenstein series representations of modular functions; see \cite{DHoker:2022dxx} for a recent review. But its physics interpretation is less clear. 

In order to understand this better, it is useful to recognize that the fermion boundary conditions (\ref{eq:bc}) can be absorbed into a background gauge field. Indeed, the effect of $(a,b)$ is to shift the levels by $(n_1, n_2) \to(n_1+a, n_2+b)$, as can be seen in (\ref{eq:eigenf}). The same spectrum is obtained if we consider fermions that are periodic on both cycles of the torus, in the presence of a flat connection\footnote{This point was also discussed before in \cite{Kim:2017xoh}.} $A_\mu$,
\be\label{eq:flat}
S= \int d^2x\, \bar \psi\left[\gamma^\mu (\partial_\mu+ i  A_\mu)- \gamma^2 \mu_F\right] \psi\,,
\ee
where
\be
 A_1=\frac{2\pi a}{L}\;,\; A_2= \frac{2\pi}{L \tau_2}(b-\tau_1 a)\,.
\ee
Therefore, the dual representation amounts to integrating over the values of this flat connection. 

Note that the gauge field in \eqref{eq:flat} is coupled to the conserved current associated to particle number. Naively, one would  conclude that this Fourier transformation corresponds to gauging the particle number $U(1)$ symmetry. However, this is not entirely correct for two reasons. On the one hand, the connection is flat and we sum over the holonomies of such a flat connection. More importantly, the transformation is done at the level of the free energy $\log Z_{ab}$ rather than the partition function $Z_{ab}$ (look for instance expression \eqref{eq:bosonic} which corresponds to a particular ${\mathbb Z}_2$ gauging of the theory). In this sense, this operation is resemblant of a disorder average, with a weight given by the Fourier factor in (\ref{eq:dualZ1}). It would be interesting to investigate whether the Fourier transformation \eqref{eq:dualZ1} has a more clear physical interpretation.

Let us now derive the dual representation for the Dirac fermion with finite chemical potential. From (\ref{eq:Zstart}), using  (\ref{eq:prod}) to perform the product over $n_2$, we have (up to the addition of a zero-point energy counterterm)
\bea
\log Z &=& \sum_{n=-\infty}^\infty \,\log \left[e^{i \pi z-i \pi \tau_1(n+a)+\pi \tau_2(n+a)}-e^{-i \pi z+i \pi \tau_1(n+a)-\pi \tau_2(n+a)} \right] + c.c. \,, 
\eea
where one fermion chirality gives the first sum, and the other chirality gives the complex conjugate term `c.c.' in this expression. The above expression can in turn be recasted in the form
\be\label{eq:Zexpmassless}
\log Z= \alpha_0 \tau_2+ \sum_{n=-\infty}^\infty \,\log \left[1-e^{-2\pi i z+ 2\pi i \tau_1(n+a)-2\pi \tau_2 |n+a|}\right]+c.c.\,.
\ee
The first term depends on the zero-point subtraction scheme. The second and third terms are rewritten as a Fourier transform with respect to the boundary conditions,
\be
\log Z = \alpha_0 \tau_2 + \sum_{l, r \in \mathbb Z}\,c_{l,r} e^{-2\pi i (l a + r b)}\;,\;c_{l,r}= \int_0^1 da \int_0^1 db\,(\log Z- \alpha_0 \tau_2) e^{2\pi i (l a + r b)}\,.
\ee

Following similar steps to those described in Appendix \ref{app:dual}, we find
\be\label{eq:Zdoublemassless}
\log Z =  \alpha_0 \tau_2 - \frac{1}{\pi} \sum_{l=-\infty}^\infty\,{\sum_{ r=-\infty}^\infty}' \frac{\tau_2}{(r \tau_2)^2+(l+r \tau_1)^2} e^{L \tau_2 \mu_F r}\,e^{-2\pi i (l a + r b)}\,.
\ee
The prime denotes that $r=0$ is not included in the sum. As we will see next, this representation gives a very efficient way of obtaining the high temperature limit.

%%%%%%%%%%%%%%%%%%%%%%%%%%%%%%%%%%%%%%%%%%%%%%%
%%%%%%%%%%%%%%%%%%%%%%%%%%%%%%%%%%%%%%%%%%%%%%%
\subsection{High energy density of states}\label{subsec:highT1}

An important observable on the torus is the energy density of states $\rho(E)$, which follows from the Laplace transform of the partition function. At low energies it is dominated by excitations near the vacuum. But at high energies it is in general quite nontrivial to compute, since it requires summing over many contributions. Cardy showed in \cite{Cardy:1986ie} that the high energy density of states for CFTs can be obtained by performing an S-dual modular transformation $\tau'=-1/\tau$. Indeed, this transformation maps high temperatures to low temperatures, and then the density of states is dominated by the vacuum in the S-dual channel. 

Our goal now is to derive the density of states at high energies and with nonzero chemical potential. We will do this in two different ways. First, we will perform an S-transformation on the partition function; see e.g. (\ref{eq:dualZ}). This follows the original steps performed by Cardy, now with $\mu_F \neq 0$. In the second approach, we will use the dual representation (\ref{eq:Zdoublemassless}). For simplicity, here we set $\tau_1=0$, while in the Appendix \ref{app:momentum} we discuss the case with nonzero momentum potential.

Consider the behavior of the partition function at high temperatures, $\tau_2 \to 0$. We focus on antiperiodic fermions, and use the S-transformation (\ref{eq:dualZ}). Recalling that
\be
\eta(i /\tau_2) \sim e^{- \frac{\pi}{12 \tau_2}}\;,\;\theta_3(\frac{L\mu_F}{2\pi}|i/\tau_2) \sim 1\;,\; \tau_2 \to 0
\ee
we have
\be\label{eq:ZSdual}
Z \sim e^{\frac{1}{2\pi}L^2 \mu_F^2 \tau_2+ \frac{\pi}{6 \tau_2}}\;,\;\tau_2 \to 0\,.
\ee
The density of states is given by the Laplace transform
\be
\rho(E) =L \int_{-i \infty}^{i \infty} d\tau_2\, e^{\tau_2 L E} Z(\tau_2)\sim \int_{-i \infty}^{i \infty} d\tau_2\, e^{\tau_2 L E+\frac{1}{2\pi}L^2 \mu_F^2 \tau_2+ \frac{\pi}{6 \tau_2}}\,.
\ee
For $\tau_2 \to 0$, this can be evaluated by saddle point approximation. The saddle point (extremum of the exponent) is
\be
\tau_2^* = \sqrt{\frac{\pi}{6} \frac{1}{LE + \frac{L^2 \mu_F^2}{2\pi}}}\,.
\ee
Hence the saddle point approximation gives
\be
\rho(E) \sim e^{2 \sqrt{\frac{\pi}{6}}\sqrt{LE + \frac{L^2 \mu_F^2}{2\pi}}}\,.
\ee
This is a Cardy-type formula in the presence of chemical potential.  Following \cite{Carlip:2000nv}, we can also compute the logarithmic correction to the density of states, obtaining
\be
\ln \rho(E) \sim 2 \sqrt{\frac{\pi}{6}\left(LE + \frac{(L\mu_F)^2}{2\pi}\right)}   - \frac{3}{4}\ln\left(LE + \frac{L^2 \mu_F^2}{2\pi}\right) \,.
\ee
The limit $\mu_F \to 0$ reproduces the CFT result (for central charge $c=1$ in this case).

We now consider a different approach, using the dual representation (\ref{eq:Zdoublemassless}). For $\tau_2 \to 0$ (and recall that $\tau_1=0$ here), the dominant contribution comes from the $l=0$ term:
\be
\log Z \approx - \frac{1}{\pi \tau_2} \,{\sum_{ r=-\infty}^\infty}'\, (-1)^r \frac{1}{r^2}e^{L \tau_2 \mu_F r}\,.
\ee
The sum with chemical potential can be found in terms of polylogarithms,
\be
\log Z \approx\,- \frac{1}{\pi \tau_2}   \left[{\rm Li}_2(-e^{L \tau_2 \mu_F})+{\rm Li}_2(-e^{-L \tau_2 \mu_F}) \right]\,.
\ee
If $\mu_F$ is fixed as $\tau_2 \to 0$, we can further expand the polylogarithms, finding
\be
\log Z \approx \frac{1}{\tau_2} \left[\frac{\pi}{6}+ \frac{1}{2\pi}(L \tau_2 \mu_F)^2 \right]\,.
\ee
This reproduces (\ref{eq:ZSdual}), now without the need of performing a modular transformation. 

The advantage of the dual representation 
(\ref{eq:Zdoublemassless}) at high temperatures is clear: the limit $\tau_2 \to 0$ is dominated by a single $l=0$ mode, while in the original expression (\ref{eq:Zexpmassless}) we would need to include at least up to $N \sim 1/\tau_2$ modes. Of course, in the present case the dual representation was not strictly required because we accessed the high temperature limit by an S-dual transformation. But we will find examples below where the dual approach is the only available option.

%%%%%%%%%%%%%%%%%%%%%%%%%%%%%%%%%%%%%%%%%%%%%%%
%%%%%%%%%%%%%%%%%%%%%%%%%%%%%%%%%%%%%%%%%%%%%%%
%%%%%%%%%%%%%%%%%%%%%%%%%%%%%%%%%%%%%%%%%%%%%%%
%%%%%%%%%%%%%%%%%%%%%%%%%%%%%%%%%%%%%%%%%%%%%%%
\section{Massive Dirac fermion}\label{sec:massive}

In this section we compute the spectrum and partition function for a massive Dirac fermion with nonzero chemical potential. This will extend the recent results of \cite{Berg:2019jhh, Downing:2023uuc, Berg:2023pca} to $\mu_F \neq 0$.

%%%%%%%%%%%%%%%%%%%%%%%%%%%%%%%%%%%%%%%%%%%%%%%
%%%%%%%%%%%%%%%%%%%%%%%%%%%%%%%%%%%%%%%%%%%%%%%
\subsection{Spectrum}

For a massive fermion, the action (\ref{eq:action0}) in the $\sigma$ variables becomes
\be\label{eq:Smassive1}
S= L \int d^2 \sigma\, \Psi^\dag \mathcal D \Psi\;,\;\Psi=\left( \begin{matrix} \psi_1 \\ \psi_2 \end{matrix}\right)\;,\;\mathcal D= \left( \begin{matrix} \bar D && L \tau_2 m \\ L \tau_2 m && D \end{matrix}\right)
\ee
and $D, \bar D$ were defined in (\ref{eq:defD}).

The eigenvalue equation
\be
\mathcal D \Psi_n = \lambda_n \Psi_n
\ee
has eigenfunctions that are plane waves, but the spinors are mixed due to the mass term
\be
\Psi= e^{2\pi i [(n_1+a) \sigma^1+(n_2+b)\sigma^2] }\left( \begin{matrix} \psi_1 \\ \psi_2 \end{matrix}\right)\,.
\ee
The eigenvalues are
\be\label{eq:lambda-massive1}
\lambda_n^\pm=2\pi i z-2\pi i \tau_1(n_1+a)+2\pi i n_2 \pm 2\pi \tau_2 \sqrt{(n_1+a)^2+ \t m^2}
\ee
where, as before, $z=b+ \frac{i}{2\pi} L \tau_2 \mu_F$, and we have introduced the dimensionless mass
\be
\t m = \frac{L m}{2\pi}\,.
\ee
The role of the chemical potential is still to complexify the thermal boundary condition $b \to z$, as in the massless case.

%%%%%%%%%%%%%%%%%%%%%%%%%%%%%%%%%%%%%%%%%%%%%%%
%%%%%%%%%%%%%%%%%%%%%%%%%%%%%%%%%%%%%%%%%%%%%%%
\subsection{Partition function}

The partition function reads
\be
Z_{ab}= \prod_{n_1, n_2} \lambda_n^+ \lambda_n^-\,.
\ee
Using the product formula (\ref{eq:prod}), the product over $n_2$ gives
\be
\prod_{n_2} \lambda_n^\pm=e^{i \pi z-i \pi \tau_1(n_1+a)\pm \pi \tau_2 \sqrt{(n_1+a)^2+ \t m^2}}-e^{-i \pi z+i \pi \tau_1(n_1+a)\mp \pi \tau_2 \sqrt{(n_1+a)^2+ \t m^2}}
\ee
The partition function then becomes (renaming $n_1 =n$)
\be
\log Z_{ab} =\alpha_0 \tau_2+ \sum_{n=-\infty}^\infty \, \log \left(1-e^{-2\pi i z+2\pi i \tau_1 (n+a)} e^{-2\pi \tau_2 \sqrt{(n+a)^2+ \t m^2}}\right)  + c.c.
\ee
As before, we will not focus on the ambiguous zero-point energy term, here denoted by $\alpha_0 \tau_2$. When $\t m \to 0$, we recover (\ref{eq:Zexpmassless}). 

Fourier-transforming with respect to the boundary conditions $(a, b)$ gives
\be\label{eq:Zinvm}
\log Z_{ab}= -2 | \t m \tau_2|  \sum_{l=-\infty}^\infty\,{\sum_{ r=-\infty}^\infty}' \,\frac{K_1\left(2\pi | \t m| |r\tau +l| \right)}{|r\tau +l | } e^{L \tau_2 \mu_F r}\,e^{-2\pi i (l a+ r b)}\,.
\ee
See App. \ref{app:dual} for more details.
Again, as a check, the limit $\t m \to 0$ agrees with (\ref{eq:Zdoublemassless}).  Furthermore, if $\mu_F=0$, this reproduces eq. (34) of \cite{Berg:2023pca}.

In Sec. \ref{subsec:modtransf} we discussed the breaking of modular invariance due to the chemical potential. The mass also breaks modular symmetry, because it is not scale invariant. Since $m^2$ is a scalar of mass dimension 2, from (\ref{eq:metric-transf}) a modular transformation maps
\be
m^2\,\to\,m'^2 = |c\tau +d|^2 m^2\,.
\ee
The representation (\ref{eq:Zinvm}) has the advantage that under a modular transformation each term retains its form (after an appropriate redefinition of $(r, l)$). 

As discussed in \cite{Berg:2023pca}, the double sum representation (\ref{eq:Zinvm}) is useful for developing conformal perturbation theory while keeping modular covariance manifest. The same will apply in our current case with nonzero chemical potential, so we will not discuss this further here. Instead, we will use these results to analyze the partition function of nonrelativistic fermions.

%%%%%%%%%%%%%%%%%%%%%%%%%%%%%%%%%%%%%%%%%%%%%%%
%%%%%%%%%%%%%%%%%%%%%%%%%%%%%%%%%%%%%%%%%%%%%%%
%%%%%%%%%%%%%%%%%%%%%%%%%%%%%%%%%%%%%%%%%%%%%%%
%%%%%%%%%%%%%%%%%%%%%%%%%%%%%%%%%%%%%%%%%%%%%%%
\section{Nonrelativistic fermions}\label{sec:nonrel}

In this last section we will analyze the nonrelativistic QFT
\be\label{eq:spinless}
S= \int dx^0 dx^1 \,\psi^\dag\left(\partial_0- \frac{1}{2m} \partial_1^2 -\mu_0 \right) \psi\,,
\ee
for a spinless fermion $\psi$. When the chemical potential vanishes, the theory is scale invariant with dynamical exponent $z=2$, namely $x^0 \to \lambda^2 x^0$ and $x^1 \to \lambda x^1$. In the presence of a chemical potential, it describes the continuum limit of a free Fermi liquid. It can arise for instance from lattice systems, and is one of the basic models in condensed matter physics \cite{Shankar:1993pf}. As we review shortly, it also describes the low energy limit of a massive Dirac fermion with chemical potential. 

Our goal will be to place this QFT on a torus, and determine its partition function. Although modular symmetry does not play an important role in this model (we cannot rotate space and time directions), we will see that the techniques developed before are still useful. In particular, they will enable us to access the high temperature limit and determine the density of states.

%%%%%%%%%%%%%%%%%%%%%%%%%%%%%%%%%%%%%%%%%%%%%%%
%%%%%%%%%%%%%%%%%%%%%%%%%%%%%%%%%%%%%%%%%%%%%%%
\subsection{Torus spectrum and nonrelativistic limit}

We put the QFT (\ref{eq:spinless}) on an euclidean torus, defined by the identifications for the spatial and thermal circles,
\be
(x^0,\,x^1) \simeq (x^0, \, x^1+L)\;,\; (x^0,\,x^1) \simeq (x^0+L \tau_2, \, x^1+L \tau_1)\,.
\ee
Before, we have denoted $x^0=x^2$.
Equivalently, in terms of the $\sigma$ variables of Appendix \ref{app:torus}, 
\be\label{eq:Snonrelsigma}
S= L \int d^2 \sigma\, \psi^\dag \left[ \partial_{\sigma^2}- \tau_1 \partial_{\sigma^1}- \frac{\tau_2}{2 L m} \partial^2_{\sigma^1}- L \tau_2 \mu_0\right] \psi\,.
\ee

As before, we consider the boundary conditions
\bea
\psi^{(ab)}(\sigma^1+1,\sigma^2)&=&e^{2\pi i a}\psi^{(ab)}(\sigma^1,\sigma^2)\,,\nonumber\\
\psi^{(ab)}(\sigma^1,\sigma^2+1)&=&e^{2\pi i b}\psi^{(ab)}(\sigma^1,\sigma^2)\,.
\eea
The eigenvalue equation
\be
%D \psi_n = \lambda_n \psi_n\;,\;D=\frac{\partial}{\partial \sigma^2}- \tau_1 \frac{\partial}{\partial \sigma^1}- \frac{\tau_2}{2 L M} \frac{\partial^2}{{\partial \sigma^1}^2}- L \tau_2 \mu
D \psi_n = \lambda_n \psi_n\;,\;D=\partial_{\sigma^2}- \tau_1 \partial_{\sigma^1}- \frac{\tau_2}{2 L m} \partial^2_{\sigma^1}- L \tau_2 \mu_0\,,
\ee
has eigenfunctions
\be
\psi= e^{2\pi i [(n_1+a) \sigma^1+(n_2+b)\sigma^2] }\,,
\ee
and eigenvalues
\be\label{eq:lambda-nonrel}
\lambda_n=2\pi i (z+n_2)-2\pi i \tau_1(n_1+a)+2\pi \tau_2 \frac{(n_1+a)^2}{2 \t m}\,,
\ee
with $\tilde m$ and $z$ defined as before.
%We have denoted
%\be
%\t m = \frac{L m}{2\pi}\;,\;z= b+ \frac{i}{2\pi} L \tau_2 \mu_0\,.
%\ee

These results can be obtained as a nonrelativistic limit of the Dirac fermion. At the level of the action, this can be done by taking a simultaneous large mass and chemical potential limits; see e.g. \cite{Daguerre:2020pte}. For our purpose, it will be useful to discuss this starting from the spectrum. This involves a large mass limit, $m L \gg 2 \pi n$, in 
(\ref{eq:Smassive1}). Simultaneously, we scale the chemical potential $\mu_F$ with the mass such that one of the eigenvalues $\lambda^\pm_n$ in (\ref{eq:lambda-massive1}) remains finite. Let us for concreteness choose
\be
\mu_F = m + \mu_0\;,\; m \gg \mu_0\,.
\ee
Then $\lambda_n^+$ reproduces the nonrelativistic eigenvalue $\lambda_n$ of (\ref{eq:lambda-nonrel}), while all the $\lambda_n^- \sim \tau_2 Lm$ are very large. We can then approximate the fermion operator as
\be
\Psi \approx \sum_n\, e^{2\pi i [(n_1+a) \sigma^1+(n_2+b)\sigma^2] } \,\psi_n \chi_n^+
\ee
where $\chi_n^+$ is the normalized spinor eigenvector for the eigenvalue $\lambda_n^+$.
The action becomes
\be
S \approx \sum_n \,\lambda_n\,\psi_n^\dag \psi_n\,.
\ee
Transforming back to position space reproduces (\ref{eq:Snonrelsigma}).

%%%%%%%%%%%%%%%%%%%%%%%%%%%%%%%%%%%%%%%%%%%%%%%
%%%%%%%%%%%%%%%%%%%%%%%%%%%%%%%%%%%%%%%%%%%%%%%
\subsection{Calculation of the partition function}

The torus partition function is given by
\be
Z_\psi(L, \tau)= \int D \psi \, D \psi^\dag\,e^{-S} \sim \text{Det}(D) \sim  \prod_n\,\lambda_n\,.
\ee
Using, as before, the product formula \eqref{eq:prod} to perform the product over Matsubara frequencies $n_2$, and renaming $n_1 = n$,
\be
Z_{ab}= \prod_n e^{i \pi z- i \pi \tau_1(n+a)+ \pi \tau_2 \frac{(n+a)^2}{2 \t m}} \left(1-e^{-2\pi iz+2\pi i \tau_1(n+a)} e^{-2\pi \tau_2 \frac{(n+a)^2}{2 \t m}} \right)\,.
\ee
Note that the factor $e^{i \pi z- i \pi \tau_1(n+a)}$ is actually cancelled if we take into account the massive modes with eigenvalues $\lambda^-_n$. This is part of the UV completion of the nonrelativistic fermion, so it is an ambiguity if we start directly from the nonrelativistic action. Let us assume that the UV completion is indeed a massive Dirac fermion. Then this factor cancels out, and
\be\label{eq:Znonrel0}
\log Z_{ab} = \sum_{n=-\infty}^{\infty}\,\log \left(1-e^{-2\pi iz+2\pi i \tau_1(n+a)} e^{-2\pi \tau_2 \frac{(n+a)^2}{2 \t m}} \right)\,,
\ee
up to a zero-point energy ambiguity.

As discussed in Sec. \ref{subsec:dual} we consider the Fourier transform with respect to the boundary conditions $(a,b)$. The calculation is described in Appendix \ref{app:dual}; it gives
\be\label{eq:Znonrel}
-\log Z_{ab}=  \sqrt{\frac{\t m}{\tau_2}} \sum_{l=-\infty}^\infty\,\sum_{r=1}^\infty\, \frac{1}{r^{3/2}}\,e^{L \tau_2 \mu_0 r}\,e^{- \frac{\pi \t m}{\tau_2 r}(l+\tau_1 r)^2}e^{-2\pi i (l a + r b)}\,.
\ee

Let us analyze how this result arises as a nonrelativistic limit of (\ref{eq:Zinvm}). Expanding the Bessel function at large $m$ gives
\be
\log Z = -\tau_2  \sum_{l=-\infty}^\infty\,{\sum_{ r=-\infty}^\infty}'  \frac{e^{-2\pi \t m \sqrt{(r \tau_1+l)^2+(r \tau_2)^2}}}{[(r \tau_1+l)^2+(r \tau_2)^2]^{3/4}} e^{-2\pi i (l a+r b)} e^{ \tau_2 L\mu_F r}\,.
\ee
Because the mass is large, all terms are exponentially suppressed, except those for which there is a near cancellation between mass and the chemical potential (tuned so that $\mu_F= m + \mu_0$, as we discussed before). This happens for the terms in the sum that satisfy
\be
(r \tau_1 +l) ^2 \ll (r \tau_2)^2\,.
\ee
Also, the near cancellation between both exponentials occurs only for one sign of $r$. So the sum over $-\infty<r<\infty$ is restricted to one sign.
Expanding the square root then reproduces (\ref{eq:Znonrel}).

%%%%%%%%%%%%%%%%%%%%%%%%%%%%%%%%%%%%%%%%%%%%%%%
%%%%%%%%%%%%%%%%%%%%%%%%%%%%%%%%%%%%%%%%%%%%%%%
\subsection{High temperatures and density of states}

The nonrelativistic theory does not seem to feature modular covariance in any useful way. But still, we have been able to find an expression for the partition function with a double sum. We will find that this is very useful for determining the high temperature limit.

Let us take the high temperature limit $\tau_2 \to 0$, $\tau_1 \sim \tau_2$. In this regime, the dominant contribution from the sum in (\ref{eq:Znonrel}) corresponds to $l=0, r=1$. Choosing also antiperiodic fermions, $a=b=1/2$, we find
\be
\log Z \approx \sqrt{\frac{\t m}{\tau_2}}\,.
\ee
Here the sign (positive) is important; it comes from $e^{2\pi i b}=-1$. So this corresponds to a negative free energy, which has the interpretation as a negative Casimir energy. 

Then the density of states follows from the integral
\be
\rho(\varepsilon) \sim\,\int d \tau_2 \, e^{\tau_2  \varepsilon + \sqrt{\frac{\t m}{\tau_2}}}\,,
\ee
where the dimensionless energy is defined as usual $\varepsilon\equiv LE$. At large $\varepsilon$ there is a saddle point at
\be
\tau_2^* =  \frac{\t m^{1/3}}{(2\varepsilon)^{2/3}}\,.
\ee
For consistency, this needs to be small --we used as input the partition function in the high temperature regime.
This is small if $\varepsilon \gg \t m^{1/2}$. If we view the model as UV completed by the massive Dirac fermion, then the $z=2$ approximation is valid for $\varepsilon \ll \t m$. So in the regime of large mass, $\t m \gg 1$, we have a parametrically big window
\be\label{eq:window}
\t m^{1/2} \ll \varepsilon \ll \t m\,,
\ee
where the $z=2$ approximation applies, and the saddle point evaluation is valid. 

Replacing the saddle point into the integral obtains
\be\label{eq:rhoCardyz2}
\rho(\varepsilon) \sim  \left(\frac{\tilde m}{\varepsilon^5}\right)^{1/6} \exp\left[\frac32 (2\t m \varepsilon)^{1/3}\right]\,.
\ee
We note the nontrivial cubic root in the exponent. So the growth in the density of states is slower than for a relativistic CFT.

This result is consistent with a scale invariant theory with dynamical exponent $z=2$. Indeed, in $d$ spatial dimensions, and with scaling $E\sim p^z$, we expect, for a local theory,
\be
\log Z \sim L^d T^{d/z}
\ee
with $L^d$ the spatial volume. We used locality for the extensivity of $\log Z \sim L^d$, and then the power of $T^{d/z}$ is fixed by scaling. Replacing this into the expression for the density of states, performing the saddle point approximation and taking $d=1$, we find
\be
\rho(\varepsilon)  \sim \exp[\varepsilon^{\frac{1}{1+z}}]\,.
\ee
See also \cite{Gonzalez:2011nz}.
For $z=2$, this reproduces the scaling we found in (\ref{eq:rhoCardyz2}). The fermion model provides an explicit and consistent derivation of an instance of this formula. In other models, it is important to keep in mind that the high energy behavior should occur below the cutoff scale for the effective description (as in (\ref{eq:window}).

%%%%%%%%%%%%%%%%%%%%%%%%%%%%%%%%%%%%%%%%%%%%%%%
%%%%%%%%%%%%%%%%%%%%%%%%%%%%%%%%%%%%%%%%%%%%%%%
%%%%%%%%%%%%%%%%%%%%%%%%%%%%%%%%%%%%%%%%%%%%%%%
%%%%%%%%%%%%%%%%%%%%%%%%%%%%%%%%%%%%%%%%%%%%%%%
\section{Future directions}\label{sec:future}

In this work we analyzed a Dirac fermion with nonzero mass and chemical potential on a two-dimensional euclidean torus. We used this paradigmatic example of a nonrelativistic QFT to understand how to generalize and modify traditional CFT methods based on modular invariance. In the three regimes that we studied in detail ($m \ll \mu_F$, $m \sim \mu_F$ and the nonrelativistic limit) we obtained explicit `dual' representations for the partition functions, by Fourier-transforming on the twisted boundary conditions $(a, b)$. The method was then used to find the high energy density of states, giving rise to generalizations of the Cardy formula to nonrelativistic models.

One could use these results to analyze different physical observables, including microcanonical entropy, finite size effects, thermal expectation values, etc. It is also interesting to note that the dual representations contain the evolution operator for a particle propagating on a torus. For instance, one can recognize in the nonrelativistic expression (\ref{eq:Znonrel}) the evolution operator for a nonrelativistic massive particle, $U(t, x) \sim \text{exp}(-m x^2/2t)$. This suggests that it could be useful to employ the worldline formalism in the present context (see e.g. \cite{Schubert:1996jj} for a review). Another direction would be to study free theories with dynamical exponent $z$ using the present methods. 

It would also be very important to consider theories in higher dimensions. In fact, CFTs in higher dimensions face problems related to the absence of modular invariance, which was also a recurrent topic in this work. Recent works have found possible generalizations of Cardy-type formulas; see e.g. \cite{Benjamin:2023qsc} and also references therein. It appears doable to extend these ideas to nonrelativistic field theory. Higher dimensions and Lorentz breaking also bring in new effects that are absent from the 2d theories we studied, such as hyperscaling violation \cite{Dong:2012se}. Lastly, it would be extremely interesting if dual representations of the type we studied could be defined for interacting theories.

\section*{Acknowledgments} 

MS and GT are supported by CONICET (PIP grant 11220200101008CO), ANPCyT (PICT 2018-2517), CNEA, and Instituto Balseiro, Universidad Nacional de Cuyo. 
JAD is a Postdoctoral Researcher of the F.R.S.-FNRS (Belgium).

\appendix

%%%%%%%%%%%%%%%%%%%%%%%%%%%%%%%%%%%%%%%%%%%%%%%
%%%%%%%%%%%%%%%%%%%%%%%%%%%%%%%%%%%%%%%%%%%%%%%
%%%%%%%%%%%%%%%%%%%%%%%%%%%%%%%%%%%%%%%%%%%%%%%
%%%%%%%%%%%%%%%%%%%%%%%%%%%%%%%%%%%%%%%%%%%%%%%
\section{Torus}\label{app:torus}

For completeness, in this Appendix we review some basic properties of the two-dimensional torus.

There are two equivalent descriptions for the torus with complex parameter $\tau=\tau_1+i \tau_2$.
First, we can consider coordinates $(x^1, x^2)$, where $x^2$ is the Euclidean time direction, and metric
\be
ds^2= (dx^1)^2+(dx^2)^2\,.
\ee
In these coordinates, the identifications are
\bea
(x^1, x^2) &\simeq & (x^1+L, x^2) \nonumber\\
(x^1,\,x^2)& \simeq &(x^1+L \tau_1, \, x^2+L \tau_2)\,.
\eea

Alternatively, we can use coordinates $(\sigma^1, \sigma^2)$ in a standard square  $0 \le \sigma^i \le 1$, i.e with identifications
\bea
(\sigma^1, \sigma^2) &\simeq & (\sigma^1+1, \sigma^2) \nonumber\\
(\sigma^1, \sigma^2) &\simeq & (\sigma^1, \sigma^2+1)\,.
\eea
In these coordinates, the metric is
\be
ds^2 = L^2 |d \sigma^1+ \tau d\sigma^2|^2\,.
\ee

The relation between the coordinates is
\be\label{eq:defw}
w= \sigma^1 + \tau \sigma^2 = \frac{1}{L}(x^1+ i x^2)\,.
\ee
In components
\be
L \sigma^1 = x^1- \frac{\tau_1}{\tau_2} x^2\;,\;L \sigma^2 =\frac{x^2}{\tau_2}\,.
\ee
For future use, we record the relations between derivatives, obtained using the chain rule:
\bea
\frac{\partial}{\partial x^1}&=& \frac{1}{L} \frac{\partial}{\partial \sigma^1} \nonumber\\
\frac{\partial}{\partial x^2}&=& \frac{1}{L \tau_2}\left(-\tau_1 \frac{\partial}{\partial \sigma^1} +\frac{\partial}{\partial \sigma^2}\right)\,.
\eea
And in terms of the complex coordinate $w$ defined in (\ref{eq:defw}),
\bea
\frac{\partial}{\partial \sigma^1}&=& \partial_w + \partial_{\bar w} \nonumber\\
\frac{\partial}{\partial \sigma^2}&=&\tau \partial_w +\bar \tau \partial_{\bar w}\,.
\eea
Then
\bea\label{eq:changedel}
\frac{\partial}{\partial x^2}+i\frac{\partial}{\partial x^1}&=&\frac{1}{L \tau_2}\left( -\bar \tau \frac{\partial}{\partial \sigma^1}+\frac{\partial}{\partial \sigma^2}\right)=\frac{2i}{L}\partial_w \nonumber\\
\frac{\partial}{\partial x^2}-i\frac{\partial}{\partial x^1}&=&\frac{1}{L \tau_2}\left( - \tau \frac{\partial}{\partial \sigma^1}+\frac{\partial}{\partial \sigma^2}\right)=-\frac{2i}{L}\partial_{\bar w}\,.
\eea
We also note that
\be
d^2x = L^2 |\tau_2|\,d^2 \sigma\,.
\ee

%%%%%%%%%%%%%%%%%%%%%%%%%%%%%%%%%%%%%%%%%%%%%%%
%%%%%%%%%%%%%%%%%%%%%%%%%%%%%%%%%%%%%%%%%%%%%%%
%%%%%%%%%%%%%%%%%%%%%%%%%%%%%%%%%%%%%%%%%%%%%%%
%%%%%%%%%%%%%%%%%%%%%%%%%%%%%%%%%%%%%%%%%%%%%%%
\section{Dual representation of the free energy}\label{app:dual}

In this Appendix we present in detail the calculation for the dual representation of the partition function. For concreteness, we focus on the nonrelativistic fermion case; the steps for a Dirac fermion are similar.

Let us denote
\be
\log Z_{ab} = \sum_{l, r \in \mathbb Z}\,c_{l,r} \,e^{-2\pi i (l a + r b)}\;,\;c_{l,r}= \int_0^1 da\,\int_0^1 db\, \log Z_{ab}\,e^{2\pi i (l a + r b)}\,.
\ee
Expanding the logarithm in (\ref{eq:Znonrel0})
\be
\log Z_{ab} = - \sum_{n=-\infty}^{\infty}\sum_{j=1}^\infty \,\frac{1}{j}\,e^{-2\pi iz j+2\pi i \tau_1(n+a)j} e^{-2\pi \tau_2 \frac{(n+a)^2}{2 \t m}j} 
\ee
we have
\be
c_{l,r}=-\sum_{n=-\infty}^{\infty} \sum_{j=1}^\infty \, \frac{1}{j}\int_0^1 da\,\int_0^1 db\,e^{-2\pi iz j+2\pi i \tau_1(n+a)j} e^{-2\pi \tau_2 \frac{(n+a)^2}{2 \t m}j}\,e^{2\pi i (l a + r b)} \,.
\ee
The integral over $b$ gives $\delta_{r,j}$, and since $j \ge 1$, this also requires $r\ge 1$. So
\be
c_{l,r}=-\theta(r-1)\sum_{n=-\infty}^{\infty}\frac{1}{r}\int_0^1 da\,e^{L \tau_2 \mu_0 r+2\pi i \tau_1(n+a)r} e^{-2\pi \tau_2 \frac{(n+a)^2}{2 \t m}r}\,e^{2\pi i l a } \,.
\ee

Next, changing variables to $a'=a+n$, and since $e^{2\pi i l a }=e^{2\pi i l a' }$ ($l$ and $n$ are integers),
\be
c_{l,r}=-\theta(r-1)\frac{1}{r}e^{L \tau_2 \mu_0 r}\,\sum_{n=-\infty}^{\infty}\int_{-n}^{1+n} da\,e^{2\pi i (l+\tau_1r) a' } e^{-2\pi \tau_2 \frac{{a'}^2}{2 \t m}r} \,.
\ee
Lastly, integrating
\bea
\sum_{n=-\infty}^{\infty}\int_{-n}^{1+n} da\,e^{2\pi i (l+\tau_1r) a' } e^{-2\pi \tau_2 \frac{{a'}^2}{2 \t m}r}&=&\int_{-\infty}^\infty\,da'\,e^{2\pi i (l+\tau_1r) a' } e^{-2\pi \tau_2 \frac{{a'}^2}{2 \t m}r}\nonumber\\
&=& \sqrt{\frac{\t m}{\tau_2 r}}\,e^{- \frac{\pi \t m}{\tau_2 r}(l+\tau_1 r)^2}
\eea
we arrive at
\be
-\log Z=  \sum_{l=-\infty}^\infty\,\sum_{r=1}^\infty\, \frac{1}{r}\sqrt{\frac{\t m}{\tau_2 r}}\,e^{L \tau_2 \mu_0 r}\,e^{- \frac{\pi \t m}{\tau_2 r}(l+\tau_1 r)^2}e^{-2\pi i (l a + r b)}\,.
\ee

%%%%%%%%%%%%%%%%%%%%%%%%%%%%%%%%%%%%%%%%%%%%%%%
%%%%%%%%%%%%%%%%%%%%%%%%%%%%%%%%%%%%%%%%%%%%%%%
%%%%%%%%%%%%%%%%%%%%%%%%%%%%%%%%%%%%%%%%%%%%%%%
%%%%%%%%%%%%%%%%%%%%%%%%%%%%%%%%%%%%%%%%%%%%%%%
\section{Density of states with nonzero momentum}\label{app:momentum}

The density of states can be computed also in the presence of nonzero $\tau_1$, which acts like a fugacity for the momentum. The high temperature limit corresponds to $\tau_2 \to 0$, $\tau_1 \sim \Omega \tau_2$, where $\Omega$ remains fixed as $\tau_2 \to 0$. Indeed, in this limit
\be
- \frac{1}{\tau}=- \frac{1}{\tau_2} \frac{\Omega-i}{\Omega^2 +1} \to \infty
\ee
and so in the modular transformed (dual) channel the vacuum gives the dominant contribution. From the definition of the Dedekind $\eta$ function, in this limit
\be
\eta(-1/\tau) \approx e^{- \frac{\pi i}{12 \tau}}
\ee
and one can check that $\theta_3 \sim 1$ as before. Then the right hand side of (\ref{eq:dualZ}) becomes
\be\label{eq:Zhigh2}
Z \approx e^{\frac{\pi i}{12 } \left(\frac{1}{\tau}-\frac{1}{\bar \tau} \right)(1+12 \t \mu^2 \tau_2^2)}\;,\;\t \mu= \frac{L \mu_F}{2\pi}\,.
\ee

In the presence of $\tau_1$, the relation between the partition function and the density of states can be obtained from
\be
Z(\tau, \bar \tau)= \sum_n e^{-\tau_2 L E_n} e^{2\pi i k_n \tau_1} \approx \int\, d\varepsilon\, dp \,\rho(\varepsilon, p)\,e^{i \frac{\varepsilon+p}{2} \tau}\,e^{-i \frac{\varepsilon-p}{2} \bar \tau}\;,\;\varepsilon= L E\;,\;p=2\pi k\,.
\ee
So we can obtain $\rho$ by Fourier transforming,
\be
\rho(\varepsilon, p) = \int d\tau d \bar \tau \,e^{-i \frac{\varepsilon+p}{2} \tau}\,e^{i \frac{\varepsilon-p}{2} \bar \tau}\,Z(\tau, \bar \tau)\,.
\ee
From (\ref{eq:Zhigh2}),
\be
\rho(\varepsilon, p) \sim \int d\tau d \bar \tau \,e^{-i \frac{\varepsilon+p}{2} \tau}\,e^{i \frac{\varepsilon-p}{2} \bar \tau}\,e^{\frac{\pi i}{12 } \left(\frac{1}{\tau}-\frac{1}{\bar \tau} \right)(1-3 \t \mu^2 (\tau- \bar \tau)^2)}\,.
\ee
Since at high energies the saddle point is at small $\tau_2$, we can self-consistently ignore the chemical potential term in order to find the saddle point. Then extremizing the exponent of the integrand and plugging the saddle value back gives
\be
\log \rho (\varepsilon, p) \approx \sqrt{\frac{\pi}{6}} \left(\sqrt{\varepsilon-p}+\sqrt{\varepsilon+p} \right) \left(1+ \frac{\pi}{4} \t \mu^2 (\frac{1}{\sqrt{\varepsilon-p}}+ \frac{1}{\sqrt{\varepsilon+p}})^2 \right)\,.
\ee

Nonzero momentum can also be incorporated in the dual representation (\ref{eq:Zdoublemassless}). For $\tau_2 \to 0$, $\tau_1= \Omega \tau_2$ (with $\Omega$ fixed) the dominant $l=0$ term gives
\be
\log Z \approx - \frac{1}{\pi} \frac{\tau_2}{\tau_1^2+\tau_2^2} \,{\sum_{ r=-\infty}^\infty}'\, (-1)^r \frac{1}{r^2}e^{L \tau_2 \mu_F r}\,.
\ee
As before, the sum can be done in terms of polylogarithms
\be
\log Z \approx\,- \frac{1}{\pi} \frac{\tau_2}{\tau_1^2+\tau_2^2} \left[{\rm Li}_2(-e^{L \tau_2 \mu_F})+{\rm Li}_2(-e^{-L \tau_2 \mu_F}) \right]\,.
\ee
The large temperature expansion at fixed $\mu_F$ reproduces (\ref{eq:Zhigh2}).

\bibliography{nonrel}{}
\bibliographystyle{utphys}

\end{document}